\documentclass[
    ,final            
  ]
 {aipproc}
\pdfoutput=1
\usepackage{multirow}
\usepackage{epstopdf}
\usepackage{epsfig}
\usepackage{amsmath}
\usepackage{units}
\usepackage{graphicx}  
\usepackage{braket}
\usepackage{xfrac}
\usepackage[normalem]{ulem}
\usepackage{booktabs}
\usepackage{tabularx}

\newcolumntype{Y}{>{\centering\arraybackslash}X}

\pdfoutput=1
\usepackage{epstopdf}
\usepackage{epsfig}
\usepackage{amsmath}
\usepackage{units}
\usepackage{graphicx}  
\usepackage{braket}
\usepackage{xfrac}
\usepackage[normalem]{ulem}
\usepackage{booktabs}
\usepackage{tabularx}
\newcolumntype{Y}{>{\centering\arraybackslash}X}

\usepackage{tabularx}
\usepackage{tabulary}
\usepackage{multirow, bigdelim}
\usepackage{bigstrut}
\usepackage{longtable}

\usepackage[table]{xcolor}
\usepackage[super]{nth}
\usepackage{enumitem}
\usepackage{hhline}
\usepackage{forloop}
\usepackage{etoolbox}
\usepackage{listings}
\usepackage[arrowdel]{physics}
\usepackage{braket}
\newcommand{\inlinecomment}[1]{}

\newcolumntype{Y}{>{\centering\arraybackslash}X}
\newcolumntype{M}[1]{>{\raggedright}m{#1}}
\newcolumntype{B}[1]{>{\noindent\raggedright}b{#1}}
\newcolumntype{P}[1]{>{\raggedright}p{#1}}

\renewcommand{\pv}{\ensuremath{PV}}

\newcommand{\N}{\ensuremath{N}}

\newcommand{\KR}[1]{KR-#1}

\renewcommand{\arraystretch}{1.25}

\layoutstyle{8x11single}

\begin{document}

\title{Development and Validation of a conceptual survey instrument to evaluate introductory physics students' understanding of  thermodynamics}


\author{Benjamin Brown}{
  address={Department of Physics and Astronomy, University of Pittsburgh, Pittsburgh, PA 15260}
}

\author{Chandralekha Singh}{
}

\begin{abstract}

We discuss the development and validation of a conceptual multiple-choice survey instrument called the Survey of Thermodynamic Processes and First and Second Laws (STPFaSL) suitable for introductory physics courses. The survey instrument uses common student difficulties with these concepts as resources in that the incorrect answers to the multiple-choice questions were guided by them. After the development and validation of the survey instrument, the final version was administered at six different  institutions.
It was administered to introductory physics students in various traditionally taught calculus-based and algebra-based classes in paper-pencil format before and after traditional lecture-based instruction in relevant concepts. 
We also administered the survey instrument to upper-level undergraduates majoring in physics and Ph.D. students for bench marking and for content validity and compared their performance with those of  introductory students for whom the survey is intended.   We find that although the survey instrument focuses on thermodynamics concepts covered in introductory courses, it is challenging even for advanced students. A comparison with the base line data on the validated survey instrument presented here
can help instructors evaluate the effectiveness of  innovative pedagogies designed to help students develop a solid grasp of these concepts. 
\end{abstract}

\date{}
\maketitle
\section{Introduction and Background}


\vspace*{-.1in}
\subsection{Multiple-choice Surveys}
\vspace*{-.1in}
Major goals of college introductory physics courses for life science, physical science and engineering majors include
helping all students develop functional understanding of physics and learn effective problem solving and reasoning skills 
  \cite{problem1,problem2,problem3,problem4,problem5,problem51,problem52,problem6,problem7,problem71,problem71,problem73}.   
Validated conceptual multiple-choice physics survey instruments administered before and after instruction in relevant concepts can be useful tools to gauge the effectiveness of curricula and pedagogies in promoting robust conceptual understanding.  
When compared to free-response problems, multiple choice problems  
can be graded efficiently and results are easier to analyze statistically for different instructional methods and/or student populations. However, multiple-choice problems also have some drawbacks. For example, students may select the correct answers with erroneous reasoning or explanation.  Also, students cannot be given partial credit for their responses.

The multiple-choice survey instruments have been used as one tool to evaluate whether research-based instructional strategies are successful in significantly improving students' conceptual understanding of  these concepts. For example, the Force Concept Inventory is a conceptual multiple-choice survey instrument that helped many instructors recognize that introductory physics students were often not developing a functional understanding of force concepts in traditionally taught courses (primarily using lectures) even if students could solve quantitative problems assigned to them by using a plug-and-chug approach \cite{fci1,fci4,fci5}. Other conceptual survey instruments at the introductory physics level in mechanics and electricity and magnetism have also been developed, including survey instruments for kinematics represented graphically \cite{beichner1994testing}, energy and momentum \cite{singh2003multiple}, rotational and rolling motion \cite{mashood1,rimoldini2005student}, electricity and magnetism \cite{ding2006evaluating,maloney2001surveying,magnetism1,magnetism2,jing2}, circuits \cite{engelhardt2004students} and Gauss's law \cite{singh2006student,gauss}. 

In thermodynamics, existing conceptual survey instruments include: 1) Heat and Temperature Conceptual Evaluation (HTCE) \cite{sokoloff,physport} that focuses on temperature, phase change, heat transfer, thermal properties of materials; 2) Thermal Concept Evaluation (TCE) \cite{shelley,shelley2} that also focuses on similar concepts to HTCE; 3) Thermal Concept Survey (TCS) \cite{manjula} that focuses on temperature, heat transfer, ideal gas law, first law of thermodynamics, phase change, and thermal properties of materials; 4) Thermodynamics Concept Inventory (TCI) \cite{tci} that focuses on concepts in engineering thermodynamics courses; and 5) Thermal and Transport Concept Inventory: Thermodynamics (TTCI:T) \cite{ttci} that also focuses on concepts in engineering thermodynamics courses. Despite these five conceptual survey instruments on introductory thermodynamics, there is a lack of research-validated survey instrument that focuses on the basic concepts related to thermodynamic processes and the first and second laws covered in introductory physics courses. Therefore, we developed  and validated \cite{engelhardt,kline1986handbook,nunnally} a 33-item conceptual multiple-choice survey instrument on these concepts called the Survey of Thermodynamic Processes and First and Second Laws (STPFaSL). We note that the overlap of the STPFaSL content with HTCE and TCE is minimal. Moreover, although there is overlap between TCI, TTCI:T and STPFaSL concepts, contexts used in TCI and TTCI:T are engineering oriented and therefore, these surveys are unlikely to be used by introductory physics instructors. Finally, TCS is for introductory physics courses and covers some common content to STPFaSL but TCS is a much broader survey and has a major emphasis on temperature, the ideal gas law, phase change and thermal properties of materials, content that are not explicitly the focus of the STPFaSL instrument.

\vspace*{-.1in}
\subsection{Inspiration from other Prior Investigations on Student Understanding of Thermodynamics}

Prior research has not only focused on the development and validation of multiple-choice surveys to investigate students' conceptual understanding of various thermodynamic concepts, but many investigations have focused on student understanding of thermodynamics without using multiple-choice surveys [38-63]. Some of these investigations use conceptual problems to probe students understanding that ask students to explain their reasoning. These investigations were invaluable in the development of the STPFaSL instrument. For brevity, below, we only give a few examples of studies that were used as a guide and from which open-ended questions were used in the earlier stages of the development of the multiple-choice questions for the STPFaSL instrument. 

 Loverude et al.\cite{loverude2002} investigated student understanding of the first law of thermodynamics in the context of how students relate work to the adiabatic compression of an ideal gas. For example, in one problem used to investigate student understanding in their study, students were asked to consider a cylindrical pump (diagram was provided) containing one mole of an ideal gas. The piston fit tightly so that no gas could escape. Students were asked to consider friction as being negligible between piston and the cylinder. The piston was thermally isolated from the surrounding. In one version of the problem, students were asked what will happen to the temperature of the gas and why if the piston is quickly pressed inward. Another type of problem posed to students in the same research involved providing a cyclic process on a PV diagram in which part of the cyclic process was isothermal, isobaric and isochoric. Students were asked whether the work done in the entire cycle was positive, negative or zero and explain their reasoning.

Another investigation of students’ reasoning of heat, work and the first law of
thermodynamics in an introductory calculus-based physics course by Meltzer et al. \cite{meltzer2004investigation} asked several conceptual problems some of which involved the PV diagrams. For example, one problem in their study involved two different processes represented on the PV diagram that started at the same point and ended at the same point. Students were asked to compare the work done by the gas and the heat absorbed by the gas in the two processes and explain their reasoning for their answers.

In another investigation focusing on student understanding of the ideal gas law using a macroscopic perspective, Kautz et al. \cite{kautz2005} asked several conceptual problems. For example, in one problem in which a diagram was provided, three identical cylinders are filled with
unknown quantities of ideal gases. The cylinders are closed with identical
frictionless pistons. Cylinders A and B are in thermal equilibrium
with the room at $20^0$C, and cylinder C is kept at a temperature of $80^0$C.
The students were asked whether the pressure of the nitrogen gas in cylinder
A is greater than, less than, or equal to the pressure of the hydrogen gas in
cylinder B, and whether the pressure of the hydrogen gas in cylinder B is
greater than, less than, or equal to the pressure of the hydrogen gas in
cylinder C. Student were asked to explain their reasoning.

Another investigation by Cochran et al. \cite{cochran2006} focused on student conceptual understanding of heat engines and the second law of thermodynamics. For example, in one question students were provided the diagram of a proposed heat engine (including the temperatures of the hot and cold reservoirs, the heat absorbed from the hot reservoir and the heat flow to the cold reservoir as well as the work done)
and asked if the device as shown could function and why.
In another investigation, Bucy et al. \cite{bucy2006entropy} focused on student understanding of entropy in the context of comparison of ideal gas processes. For example, students were asked to compare the change in entropy of an ideal gas in an isothermal expansion and free expansion into a vacuum and also explain whether the change in entropy of the gas in each case is positive, negative or zero in each case and why.
In another investigation by Christensen et al. \cite{christensen2009student}, students' ideas regarding entropy and the second law of thermodynamics in an introductory physics course were studied. They found that students struggled in distinguishing between entropy of the system and the surrounding and had great difficulty with spontaneous processes. 
Another investigation by Smith et al. \cite{john2} focused on student difficulties with concepts related to entropy, heat engines and the Carnot Cycle and how student understanding can be improved. 

\vspace*{-.1in}
 \subsection{Goal of this paper} 
Here we discuss the development and validation of the STPFaSL instrument related to thermodynamic processes and the first and second laws covered in introductory physics courses in which these prior investigations were used as a guide. 
We present average base line data from the STPFaSL survey  instrument from traditional lecture-based introductory physics courses (along with the STPFaSL survey instrument and key in Ref. [\cite{epaps}]) so that instructors in courses covering the same concepts but using innovative pedagogies can compare their students' performance with those provided here to gauge the relative effectiveness of their instructional design and approach. The data were collected from six different higher education institutions in the US (four research-intensive large state universities and two colleges). Since the data from different institutions for the same type of course (e.g., calculus-based introductory physics course) are similar, average combined data from different institutions for the same course type are presented.
  \vspace*{-.1in}
 \section{STPFaSL Instrument Development and Validation} 
\vspace*{-.1in}
 The thermodynamics survey instrument development and validation process was analogous to those for the earlier conceptual survey instruments developed by our group \cite{singh2003multiple,rimoldini2005student,magnetism1,magnetism2,singh2006student}. Our process is consistent with the established guidelines for test development and validation \cite{engelhardt,kline1986handbook,nunnally} using the Classical Test Theory (CTT).  According to the standards for the multiple-choice test (survey) instrument design, a high-quality test instrument  
 has five characteristics: reliability, validity, discrimination, good comparative data and suitability for the population \cite{engelhardt,kline1986handbook,nunnally}. Moreover, the development and validation of a well-designed survey instrument is an iterative process that should involve recognizing the need for the survey instrument, formulating the test objectives and scope for measurement, constructing the test items, performing content validity and reliability check, and distribution \cite{engelhardt,kline1986handbook,nunnally}. Below we describe the development and validation of the STPFaSL instrument.
 
 \subsection{Development of Test Blueprint}

Before developing the STPFaSL instrument items, we first developed a test blueprint to provide a framework for deciding the desired test attributes. The test blueprint provided an outline and guided the development of the test items. The development of the test blueprint entailed formulating the need for the survey instrument, determining its scope, format and testing time of the test as well as determining the weights of different sub-topics consistent with the scope and objective of the test. The specificity of the test plan helped to determine the extent of content covered and the complexity of the questions. 

As noted in the introduction, despite the existence of several thermodynamics survey instruments at the introductory level \cite{sokoloff,physport,shelley,shelley2,manjula,tci,ttci}, there is no research-validated survey instrument that focuses on the basics of thermodynamic processes and the first and second laws of thermodynamics covered in the introductory physics courses. Therefore, we developed and validated the STPFaSL instrument focusing on content covered in introductory physics courses. The STPFaSL instrument is a multiple-choice conceptual survey on thermodynamic processes and the first and second laws covered in both calculus-based and algebra-based introductory physics courses. It can be used to measure the effectiveness of traditional and/or research-based approaches for helping introductory students learn thermodynamics concepts covered in the survey for a group of students.  Specifically, the survey instrument is designed to be a low stakes test to measure the effectiveness of instruction in helping students in a particular course develop a good grasp of the concepts covered and is not appropriate for high stakes testing. The STPFaSL survey instrument can be administered before and after instruction in relevant concepts to evaluate introductory physics students' understanding of these concepts and to evaluate whether innovative curricula and pedagogies are effective in reducing the difficulties. With regard to the testing time, this survey is designed to be administered in one 40-50 minute long class period although instructors should feel free to give extra time to their students as they deem appropriate. The survey can also be administered in small groups in which students can discuss the answers with each other before selecting an answer for each item. With regard to the weights of different sub-topics consistent with the scope and objective of the test, we browsed over introductory physics textbooks, consulted with seven faculty members and looked at the kinds of questions they asked their students in homework, quizzes and exams before determining it as discussed below.

  \subsection{Formulating Test Objectives and Scope} 
 
We focused the survey content on thermodynamic processes and first and second laws that is basic enough that the survey instrument is appropriate for both algebra-based and calculus-based introductory physics courses in which these thermodynamics topics are covered.  We also made sure that the  survey instrument has questions at different levels of cognitive achievement \cite{engelhardt,kline1986handbook,nunnally}.

In order to formulate test objectives and scope pertaining to thermodynamic processes and first and second laws, the survey instrument development started by consulting with seven instructors who regularly teach calculus-based and algebra-based introductory physics courses in which these topics in thermodynamics are covered. We asked them about the goals and objectives they have when teaching these topics and what they want their students to be able to do after instruction in relevant concepts. In addition to perusing through the coverage of these topics in several algebra-based and calculus-based introductory physics textbooks, we browsed over homework, quiz and exam problems that these instructors in introductory algebra-based and calculus-based courses at the University of Pittsburgh (Pitt) had typically given to their students in the past before determining the test objective and scope of the test in terms of the actual content and starting the design of the questions for the instrument. The preliminary distribution of questions from various topics was discussed, and iterated several times and finally agreed upon with seven introductory physics course instructors at Pitt.

 \begin{table}
 	
 	\setlength{\tabcolsep}{0pt}
 	\renewcommand{\arraystretch}{0.95}
 	\newlength{\columnone}
 	\setlength{\columnone}{3.5cm}
 	\caption{Topics by item number.  Pitt physics faculty members and the Pitt PER group independently reached a consensus on identifying topics involved in each problem.   A ``M'' indicates that a concept was mentioned but not required to solve the problem in the opinion of content experts.  A ``R'' indicates that a concept is required to solve the problem.  An ``I'' indicates that a topic is implicitly required though not explicitly asked for or mentioned.  For instance, generally, if a student must reason about a \pv diagram to infer the heat transfer to a system, the concept of ``work'' is implicitly required.}
 	\label{TopicsbyItem}
 	\resizebox{0.97\columnwidth}{!}{
 		
 		\begin{tabular}
 			{>{\bfseries}m{\columnone} l | >{}cc>{}cc>{}cc>{}cc>{}cc>{}cc>{}cc>{}cc>{}cc>{}cc>{}cc>{}cc>{}cc>{}cc>{}cc>{}cc>{}c|}
 			
 			\hhline{*{2}~*{33}-}
 			
 			{} & {}  & 1 & 2 & 3 & 4 & 5 & 6 & 7 & 8 & 9 & 10 & 11 & 12 & 13 & 14 & 15 & 16 & 17 & 18 & 19 & 20 & 21 & 22 & 23 & 24 & 25 & 26 & 27 & 28 & 29 & 30 & 31 & 32 & 33 \\
 			\hline
 			\multirow{7}{\columnone}{Processes}
 			& Reversible & R & R & R & R & & & & & & & & & & & R & R & & & R & & & & & & & & & R&&&&&\\
 			& Irreversible & & & & & & & & & & & R & R & R & R & & R & & & & & & & R & R & R & & & & R & & & R & R \\
 			& Cyclic & & & & & R & R & R & R & & & & & & & R & R & & R & R & & & & & & & & & & & & & &\\
 			& Isothermal & & & R & R & & & & & R & & & & & & & & & R & & R & & & & & & M & R & R & & & & &\\
 			& Isobaric & & & & & & & & & R & & & & & & & & M & & & R & R & & & & & & & & & & & &\\
 			& Isochoric & & & & & & & & & & & & & & & & & M & & & & R & & & & & & & & R & R & & &\\
 			& Adiabatic & R & R & & & & & & & & & & & & & & & & R & & R & & & & & & M & & & & R & & &\\
 			\hline
 			\hline
 			\multirow{3}{\columnone}{Systems} & Systems \& Universe & R & & & & & & & & & & R & R & R & R & & & & & R & & & & R & R & R & & & R & & & R & R & \\
 			& Isolated System & & & & & & & & & & & R & R & & R & & R & & & & & & & R & R & & & & & & & R & R & \\
 			& Ideal Gas & & & R & R & & & & & & & & & & & & & R & & & & & & & & & & R & R & & R & R & R & \\
 			\hline
 			\hline
 			\multirow{8}{\columnone}{Quantities \& Relations} & State Variables & & & & & R & R & R & R & & R & & & & & R & & & & & & & R & & & & & & & & & & &\\
 			\hhline{*{2}~*{33}-}
 			& Internal Energy & & R & I & & R & & & & I & & R & & & R & R & & I & & & & R & R & R & & & & I & I & R & I & R & &\\
 			& \quad  Relation to T & & R & R & & & & & & R & & & & & & & & R & & & & & & & & & & R & R & & R & & &\\
 			\hhline{*{2}~*{33}-}
 			& Heat & & & & R & & & & R & R & R & & & R & & R & & & R & R & & R & & & & & & R & & & & & &\\
 			\hhline{*{2}~*{33}-}
 			& Work & & I & R & & & I & & R & I & I & & & & I & I & & & & & & I & R & R & & & I & I & & I & R & I & &\\
 			& \quad Relation to p, V & & R & & & & R & & & R & R & & & & R & R & & & & & & R & & R & & & R & R & & R & R & R & &\\
 			\hhline{*{2}~*{33}-}
 			& Entropy & R & & & R & & & R & & & & & R & & R & R & R & & & R & & & R & & R & & & & R & R & & & R &\\
 			& \quad Relation to Q, T & R & & & R & & & & & & & & R & & & & & & & R & & & & & R & & & & R & R & & & &\\
 			\hline
 			\hline
 			Representation & \pv Diagram & & & & & R & R & R & R & R & R & & & & & R & & R & & & & R & & & & & R & & & & R & & &\\
 			\hline
 			\hline
 			First Law & 1st Law & & R & R & & & & & R & R & R & R & & R & R & R & & & & & & R & & R & & R & & R & & R & R & R & & R \\
 			\hline
 			\hline
 			\multirow{2}{\columnone}{Second Law} & 2nd Law & & & & & & & & & & & R & R & R & R & & R & & & R & & & & & R & R & & & R & R & & & R & R \\
 			&Engine Efficiency & & & & & & & & & & & & & & & & R & & & R & & & & & & & & & & & & & &\\
 			\hline
 		\end{tabular}
 	}
 \end{table}
 \subsection{Concepts Covered}

 Table ~\ref{TopicsbyItem} shows that the broad categories of topics covered in the survey are {Processes,} {Systems,} {Quantities \& Relations,} {Representation,} the {First Law of Thermodynamics,} and the {Second Law of Thermodynamics}.  The Processes category includes items which require understanding of thermodynamic constraints such as whether a process is reversible, isothermal, isobaric or adiabatic. Also included are problems involving irreversible and cyclic processes.  We note that these different processes are not necessarily exclusive, e.g., one can consider an isothermal reversible process. The Systems category includes items involving knowledge of the distinction between a system and the universe, items involving subsystems or an isolated system.  The Systems category also includes items in which a student could make progress by making use of the fact that the system is an ideal gas (e.g., for an ideal gas, the internal energy and temperature have a simple relationship which can be used to solve a problem).   Quantities and Relations includes survey items specific to a quantity such as internal energy, work, heat, entropy, and their quantitative relationships.  For example, the relationship between work and the area under the curve on a \pv diagram is tested in several problems.  The Representation category includes items in which a process is represented on a \pv diagram.  Finally, the last two categories include items requiring the first law and second law of thermodynamics. We classified questions about heat engines into the Second Law of Thermodynamics category (although heat engines involve both the first and second laws) due to the particular focus of the only two problems on the survey that touched upon heat engines.

 \subsection{Development of 
 	Multiple-Choice Test Items} 
 
 As noted, the selection of topics for the questions included consultation with 7 instructors who teach introductory thermodynamics (some of whom had also taught upper-level thermodynamics) about their goals and objectives and the types of conceptual and quantitative problems they expected their students in introductory physics courses to be able to solve after instruction.  The wording of the questions took advantage of the existing literature regarding student difficulties in thermodynamics, input from students' written responses and interviews and input from physics instructors who teach these topics.  
 However, most questions on the survey require reasoning, and there are very few questions 
 which can be answered simply by rote memory. 
 
  Since we wanted instructors to be able to administer STPFaSL instrument in one 40-50 minute long class period, the final version of the survey has 33 multiple-choice items (see Ref.[\cite{epaps}]). Each question has one correct choice and four alternative or incorrect choices. We find that most students are able to complete the survey in one class period. We note however that instructors can choose to give longer time to their students as they see fit. 
  
  In developing good alternative choices for the multiple-choice conceptual problems, we first took advantage of prior work on student difficulties with relevant topics in thermodynamics [38-63]. To investigate student difficulties further in introductory physics courses at University of Pittsburgh (Pitt), we administered sets of free-response questions to students in various introductory physics courses after traditional instruction in which students had to provide their reasoning. Many of these questions were similar to the type of open-ended conceptual free-response problems that were summarized in introductory section from prior studies. While many of the findings replicated what was found in prior investigations, the responses to these open-ended questions were summarized and categorized to understand the prevalence of various difficulties at Pitt. These findings will be presented in future publications. In addition to leveraging the findings of prior research on students' conceptual understanding of these concepts [38-63], the process of administering some open-ended questions at Pitt was helpful in order to internalize the findings of prior research and develop good alternatives for the questions in the survey based upon common difficulties.

 Moreover, as part of the development and validation of the survey,   T
 the concepts involved in the STPFaSL instrument and the wording of the questions have been independently evaluated by four physics faculty members who regularly teach thermodynamics at Pitt (in addition to the feedback from members of the Physics Education Research or PER group at Pitt) and iterated many times until agreed upon.  Moreover, two faculty members from other universities who are experts in thermodynamics PER provided invaluable feedback several times to improve the quality of the survey questions.

\subsection{Refinement of Test Items based upon Student Interviews}

We also interviewed individual students using a think-aloud protocol at various stages of the survey instrument development to develop a better understanding of students' reasoning processes when they were answering the free-response and multiple-choice questions. Within this interview protocol, students were asked to talk aloud while they answered the questions so that the interviewer could understand their thought processes. Individual interviews with students during development of the survey instrument were useful for an in-depth understanding of the mechanisms underlying common student difficulties and to ensure that students interpreted the questions appropriately. Based upon the student feedback, the questions were refined and tweaked.

We note that during the initial stage of the development and validation process, 15 students in various algebra-based and calculus-based physics courses  participated in the think-aloud interviews.  Ten graduate students and undergraduates who had learned these concepts in an upper-level thermodynamics and statistical mechanics course were also interviewed. The purpose of involving some advanced students in these interviews was to compare the thought processes and difficulties of the advanced students in these courses with introductory students for bench marking purposes. This type of bench marking has been valuable to illustrate growth of student understanding in prior research \cite{bench}. We found that students' reasoning difficulties across different levels are remarkably similar except in a few instances, e.g., advanced students were more facile at reasoning with PV diagrams than introductory students. Moreover, nine additional interviews, drawn from a pool of students in introductory courses who were currently enrolled in a second semester course after finishing the first semester course (in which mechanics and thermodynamics were covered), were conducted with the STPFaSL instrument when it was close to its final form to tweak the wording of the questions further.  

 \subsection{Refinement of Test Items based upon Instructor Feedback}
 We note that in addition to developing good distractors by giving free-response questions to students and interviewing students with different versions of the multiple-choice survey, ongoing expert feedback was essential. We not only consulted with faculty members initially before the development of the survey questions, but also iterated different versions of the open-ended and multiple-choice questions with several instructors at Pitt at various stages of the development of the survey. Four instructors at Pitt reviewed the different versions of the STPFaSL instrument several times to examine its appropriateness and relevance for introductory algebra-based or calculus-based courses and to detect any possible ambiguity in item wording. Also, as noted, two faculty members 
  from other universities who have been extensively involved in physics education research in thermodynamics also provided extremely valuable suggestions and feedback to fine-tune the multiple-choice version many times into the final form.  
  
 \subsection{Fine-tuning of the Survey based upon Statistical Analysis} 
  
\begin{figure}
 	\includegraphics[width=2.2in]{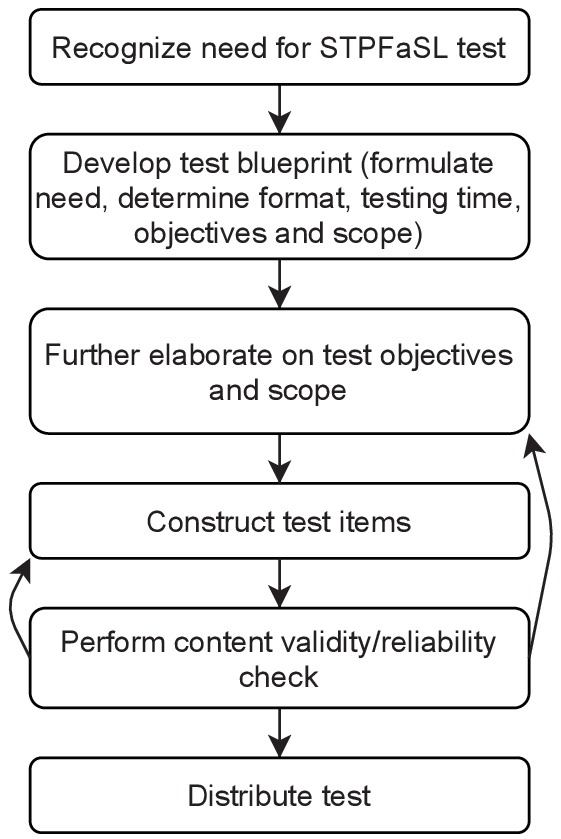}
 
 	\caption { A schematic diagram of the STPFaSL instrument development process.
 	}
 	\label{diagram1}
 	
 \end{figure}

   On the STPFaSL instrument, the incorrect choices for each item often reflect students' common alternative conceptions to increase the discriminating properties of the item. Having good distractors as alternative choices is important so that the students do not select the correct answer for the wrong reason. Statistical analysis based upon classical test theory (to be discussed in the next section) was conducted on different versions of the multiple-choice survey instrument as the items were being refined, which helped fine-tune the items further. A schematic diagram of the STPFaSL instrument development process is shown in Figure \ref{diagram1}.

\subsection{Students' Knowledge of Survey Content Before Introductory Physics} 

Discussions with students suggested that introductory physics students had some knowledge of thermodynamics from high school physics and chemistry courses, college chemistry courses and/or medical school entrance exam preparatory materials. Therefore, although a majority of the open-ended questions were administered after traditional instruction in relevant concepts, we wanted to gain some insight into what introductory physics students knew about the relevant thermodynamic concepts in the survey instrument from previous courses before they learned about them in that course. 
Therefore, we administered the following brief open-ended survey as bonus questions in a midterm exam (for which students obtained extra credit) to students in the first semester of an algebra-based physics course in which the instructor had started discussing thermodynamics, introducing concepts such as temperature, heat capacity, thermal expansion and heat transfer, but there was no instruction in the first and second laws of thermodynamics:
\begin{enumerate}
	\item Describe the first law of thermodynamics in your own words.
	\item Describe the second law of thermodynamics in your own words.
	\item Describe other central or foundational principles of thermodynamics (other than the first and second laws).
\end{enumerate}
Of the 207 students, 134 chose to respond to at least some of these bonus questions (65\%).  Their responses about the laws of thermodynamics and the difficulties they reveal are shown in Table ~\ref{algebrapreprinciples}. In particular, 
we find that for the first law question, while about half of the students stated that energy is conserved, e.g.,``Energy cannot be created or destroyed'' (52\%), only 5\% made a statement that includes heat transfer as part of the conservation law.  Another frequent response to the first law question was that heat itself is conserved, with 15\% of students making statements such as ``There is no loss of total heat in the universe.'' These responses  
confirmed that many students in introductory physics have been exposed to the first and second laws of thermodynamics before instruction in the college physics course and the survey can be administered as a pre-test before instruction in introductory courses.

\begin{table}
	\caption{Responses of introductory physics students in an algebra-based course about the laws of thermodynamics based upon what they had learned in previous courses before instruction in these laws in that physics course. The percentages are determined by taking into account only the students who attempted to answer the bonus questions.}
	\label{algebrapreprinciples}%
	\begin{tabular*}{0.70\linewidth}{>{\bfseries}llr}
		Topic                                            & Claim                     & Frequency (\%)\\ 
		\hline
		{\multirow{4}{*}{First Law}} & Energy is conserved (no mention of heat)      & 52      \\
		{}                           & Energy is conserved, with heat somehow incorporated & 5       \\
		{}                           & Heat is conserved         &15           \\
		\cline{2-3}
		{}                           & Total first law-like responses         &72           \\
		\hline
		{\multirow{5}{*}{Second Law}} & Entropy increases always               & 14\\
		{}                   & Entropy increases under some conditions                          &  4         \\
		{}                   & Energy becomes unusable                           &   10    \\
		{}                   & Heat flows from warmer objects to cooler objects & 23          \\
		\cline{2-3}
		{}                           & Total second law-like responses         &51       \\
		\hline
		
	\end{tabular*}
\end{table}

\section{Validation of the Survey Instrument} 

While developing and validating the STPFaSL instrument, we paid particular attention to the issues of reliability and validity.  Test reliability refers to the relative degree of consistency between the scores if an individual immediately repeats the test, and validity refers to the appropriateness of interpreting the test scores. We note that the STPFaSL instrument is appropriate for making interpretations about the effectiveness of instruction in relevant concepts in a particular course and it is not supposed to be used for high stakes testing of individual students. Also, although the survey instrument focuses on concepts that are typically covered in introductory thermodynamics and is appropriate for introductory students in physics courses, it was also validated and administered to undergraduates in upper-level thermodynamics and statistical mechanics courses in which these concepts are generally repeated and to first year physics Ph.D. students in order to obtain base line data and to ensure content validity (on average, advanced students should perform better than the introductory students for content validity).

Below, we describe the STPFaSL instrument in terms of the quantitative measures used in the classical test theory for a reliable survey instrument including item analysis (using item difficulty and point biserial coefficient) and  KR-20 
  \cite{engelhardt,kline1986handbook}. We also discuss the content validity of the STPFaSL survey instrument using comparison with advanced student performance (the fact that the advanced students performed significantly better than introductory physics students on the instrument) and the stability of the introductory physics student responses when the order of distractors is switched in each item.

\subsection{Overall Performance and Item Difficulty} 
 
Table 3 shows the number of students in each group to whom the final version of the survey was administered as well as the average performance of different groups on the entire survey instrument and on subsets of items focusing on particular themes.  In this Table, the average data from six 
institutions are presented because there was no statistically significant difference between the scores. In introductory courses, the pretest was administered before students learned about thermodynamics in that course and the posttest was administered after instruction in relevant concepts.  The instructors generally administered pretests in their classes by awarding students bonus points as incentive to take the survey seriously but generally awarded students a small amount of quiz grade for taking it as a posttest. Moreover, since thermodynamics is covered after mechanics in the same course, some instructors teach it at the end of the first semester introductory physics course while others teach it at the beginning or in the middle of a second semester course. Furthermore, some instructors only spent two weeks on these topics whereas others spent up to four weeks in the introductory courses. However, we find that the scores in introductory courses were not statistically significantly different across the same type of course (algebra-based or calculus-based introductory physics course) taught by different instructors in different institutions regardless of the duration over which these topics were discussed. This may at least partly be due to the fact that students in the introductory physics courses in general performed very poorly on the posttest after traditional instruction (see Table 3). Table 3 also lists Hake normalized gain $g$ defined as $g=(post\%-pre\%)/(100\%-pre\%)$ for introductory courses \cite{fci4} for which both pretest and posttest data are available. The normalized gains show that introductory physics students did not improve much from pretest to posttest.

The item difficulty of each multiple-choice question on the instrument is simply the percent of students who correctly answered the question, i.e., it is the
   average score on a particular item.  
Results in the Table 4 show not only the item difficulty of each question on the instrument but also the prevalence of different incorrect choices for each question for each group. 

\subsection{Point Biserial Coefficient} 

\begin{figure}

\includegraphics[width=1.8in]{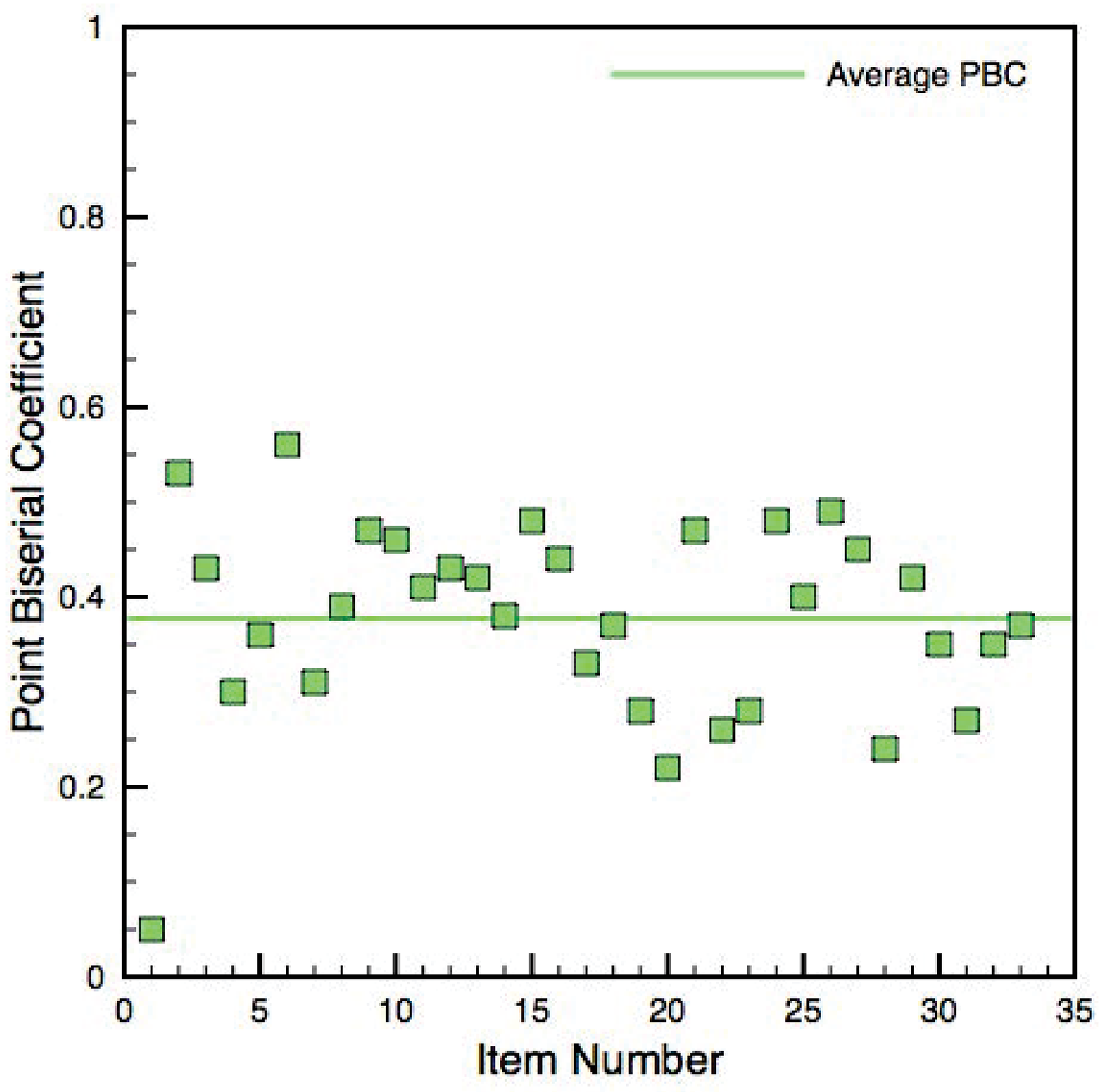}
\caption { PBC or point biserial coefficient for each item. The line in the figure presents the mean value for all items.
}
\label{upperperformancepbc}

\end{figure}

    The Point Biserial Coefficient, or PBC, is designed to measure how well a given item predicts the overall score on a test.  It is defined as the correlation coefficient between the score for a given item and the overall score.  
    The PBC can take on values between -1 and 1; a negative value indicates that otherwise high-performing students score poorly on this item, and otherwise poorly-performing students do well on the item.    
    The point biserial coefficients are shown in Figure \ref{upperperformancepbc}. A widely used criterion \cite{kline1986handbook} is that it is desirable for this measure to be greater than or equal to 0.2,
    which is exceeded for 32 of the 33 items on the STPFaSL.  The first item, which was considered to be a valuable item by experts (and hence is kept), has low PBC due to the fact that  even those students who perform well overall have difficulty distinguishing whether the change in the entropy of a system in a reversible adiabatic process is zero because the reversible process is adiabatic or whether the change in entropy of the system is zero in reversible processes in general (partly due to confusion between the system and the universe). 
   
\begin{table}[tb]
  \caption{The average performance of different groups on all of the 33 items taken together or a subset of items and the number (N) of students who participated in the survey in each group.  ``Upper-under'' consists of advanced undergraduate students who had learned the relevant concepts in an upper-level thermodynamics and statistical mechanics course, ``Ph.D. student Ind." (where ind. stands for individual) consists of entering physics Ph.D. students in their first semester of the Ph.D. program.  ``Ph.D. student pairs" consist of small groups (20 pairs and one group with 3) of Ph.D. students discussing and responding to the survey together.  Pretest (pre) was administered at the beginning of the course and posttest (post) was administered at the end of the course in the introductory physics courses. The normalized gain, g, is listed for introductory courses for which both pre/post data are available. Instructors' data are not shown here, as those data cannot be considered in a statistical manner. Four instructors self-reported that they performed near-perfect, missing zero to two items.  
  }
\label{tablesummary}
  \centering

  \begin{tabularx}{0.95\linewidth}{l | X X X X X X X}
  &Ph.D.\hspace{5mm} student & Ph.D.\hspace{5mm} student & Upper-under & Calculus-based & Calculus-based & Algebra-based & Algebra-based \\
  &Pairs & Ind. & Post & Pre & Post (g) & Pre & Post (g) \\

  \hline
  $N$ &21& 45& 147& 705& 507& 218& 382\\
  \hline
  Total Score (\%) & 75& 55& 57& 29& 37 (0.11)& 30& 37 (0.10)\\
  First Law (\%)& 76& 58& 60& 29& 37 (0.11)& 28& 38 (0.14)\\
  Second Law (\%)& 74& 56& 60& 28& 36 (0.11)& 29& 42 (0.18) \\
  PV Diagram (\%)& 71& 53& 56& 28& 38 (0.14)& 29& 29 (0.0) \\
  Reversible (\%)& 65& 44& 38& 22& 27 (0.06)& 22& 27 (0.06)\\
  Irreversible (\%)& 79& 62& 66& 32& 40 (0.12)& 32& 45 (0.19)\\
  \hline
  \end{tabularx}
\end{table}

\subsection{Reliability} 

One way to measure reliability of the test instrument is to prepare an ensemble of identical students, administer the test instrument to them, and analyze the resulting distribution of item and overall scores.  Since this is generally impractical, instead, a method is devised to use subsets of the test itself, and consider the correlation between different subsets.  The Kuder-Richardson reliability index or \KR{20} reliability index \cite{engelhardt,kline1986handbook,nunnally}, which is a measure of the self-consistency of the entire test instrument, can take a value between 0 and 1 (it divides the full instrument into subsets and the 
consistency between the scores on different subsets is estimated).
If guessing is high, KR-20 will be low. 
The KR-20 for introductory calculus-based and algebra-based courses was 0.77 and 0.61, respectively, after instruction and for graduate students and upper-level undergraduates used to bench mark the survey was 0.87 and 0.79, respectively. These values are reasonable for predictions about a group of students and the higher values for students with more robust knowledge are expected due to lower guessing \cite{engelhardt,kline1986handbook,nunnally}.

\subsection{Content Validity via Administration to Students Groups at Different Levels} 
 
   The survey instrument administration to upper-level students and Ph.D. students is useful for content validity.  
The content validity refers to the fact that the performance of students on the survey instrument closely corresponds to the model of expected performance. One measure of content validity can come from the expectation that introductory students will be out-performed by upper-level undergraduates and Ph.D. students and  pairs of Ph.D. students working together will outperform those working individually. 
 More than a thousand students from introductory courses in which these concepts have been covered have been administered the final version of the survey instrument and upper level undergraduates in thermodynamic and statistical mechanics courses have participated for the purposes of bench marking and content validity of the instrument (see Table \ref{tablesummary}). In addition,  
 45 entering physics Ph.D. students in their first semester of the Ph.D. program (who had not yet taken Ph.D. level thermodynamics) were administered the survey instrument individually and in pairs (after working on it individually).  
  
  The mean performance across the 33 items and the number of students who participated in the survey at each level are shown in Table ~\ref{tablesummary}.  ``Upper-under'' consists of advanced undergraduates and ``Ph.D. students Ind." refers to entering Ph.D. students in their first semester of the Ph.D. program taking the survey individually.  Ph.D. student pairs consist of small groups (most in pairs and one group with 3) of Ph.D. students discussing and responding to the survey together.  Moreover, four faculty members who teach thermodynamics regularly and took the survey self-reported that they performed nearly perfectly, missing zero to two items. Thus, as one considers levels from advanced to introductory, performance deteriorates.  From pairs of Ph.D. students to pretest scores for introductory students, the average performance drops from 75\% to 29\%.  The average data for each group tabulated in Table~\ref{tablesummary} and the expected trends observed serve as a measure of content validity.
 
  \subsection{Effect of Ordering Distractors on Student Performance} 
  
  We performed an investigation to evaluate a different form of reliability and validity of the STPFaSL instrument.  In particular, the answer choices were re-ordered to determine if answer choice ordering had an effect on student performance.  One version was the original order, and three more versions differing only in answer choice order were administered to students in a calculus-based introductory physics course after instruction in relevant concepts.  Students were randomly assigned one of these versions. Performing a Kruskal-Wallis test for statistically significant difference between any of the four sets found no difference. In particular,  the p-value that differences between the four groups were due to chance alone was found to be 0.994 \cite{kline1986handbook}. This study was performed with a total of 226 students who scored an average of 36.9\%, which was typical of the average performance of students in a calculus-based introductory course after instruction from different universities (posttest) (see Table\ref{tablesummary}).

\subsection{A Glance to Student Difficulties on the Validated Survey} 
  
\begin{figure}[htbp]

\includegraphics[width=\textwidth]{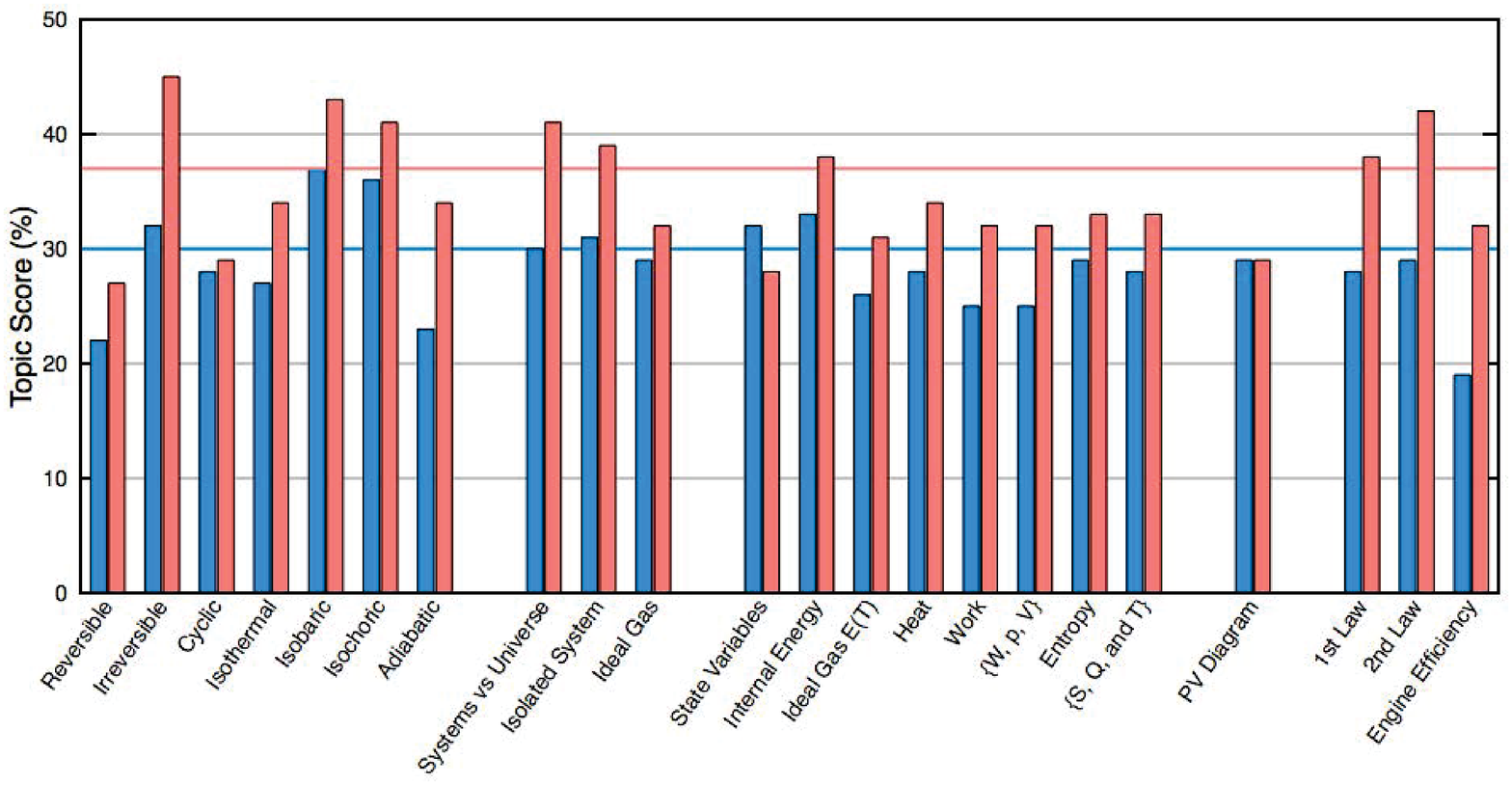}
\caption{Average percentage scores for students in algebra-based introductory physics courses on the STPFaSL instrument by topic before and after instruction (for the pretest, the number of students \N=218, blue, and for the posttest, \N=382, red).  
 The blue and red horizontal lines show the averages on the entire survey instrument before and after instruction, respectively.
}
\label{TopicScoresAlgebra}

\end{figure}

\begin{figure}[htbp]

\includegraphics[width=\textwidth]{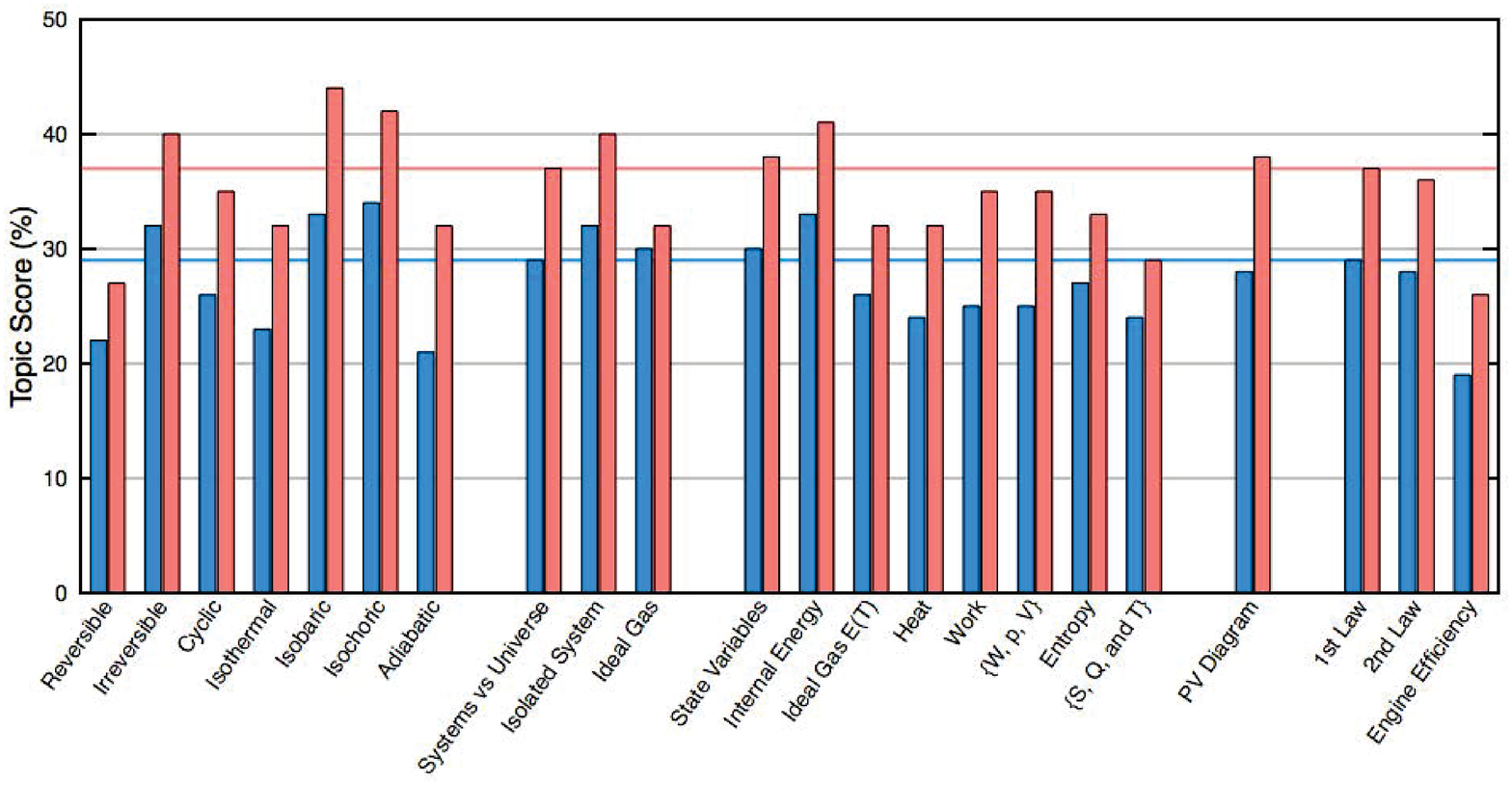}
\caption{Average percentage scores for students in calculus-based introductory physics courses on the STPFaSL instrument by topic before and after instruction (for the pretest, the number of students \N=704, blue, and for the posttest, \N=505, red). The blue and red horizontal lines show the averages on the entire survey instrument before and after instruction, respectively.}
\label{TopicScoresCalculus}

\end{figure}

\begin{figure}[htbp]

\includegraphics[width=\textwidth]{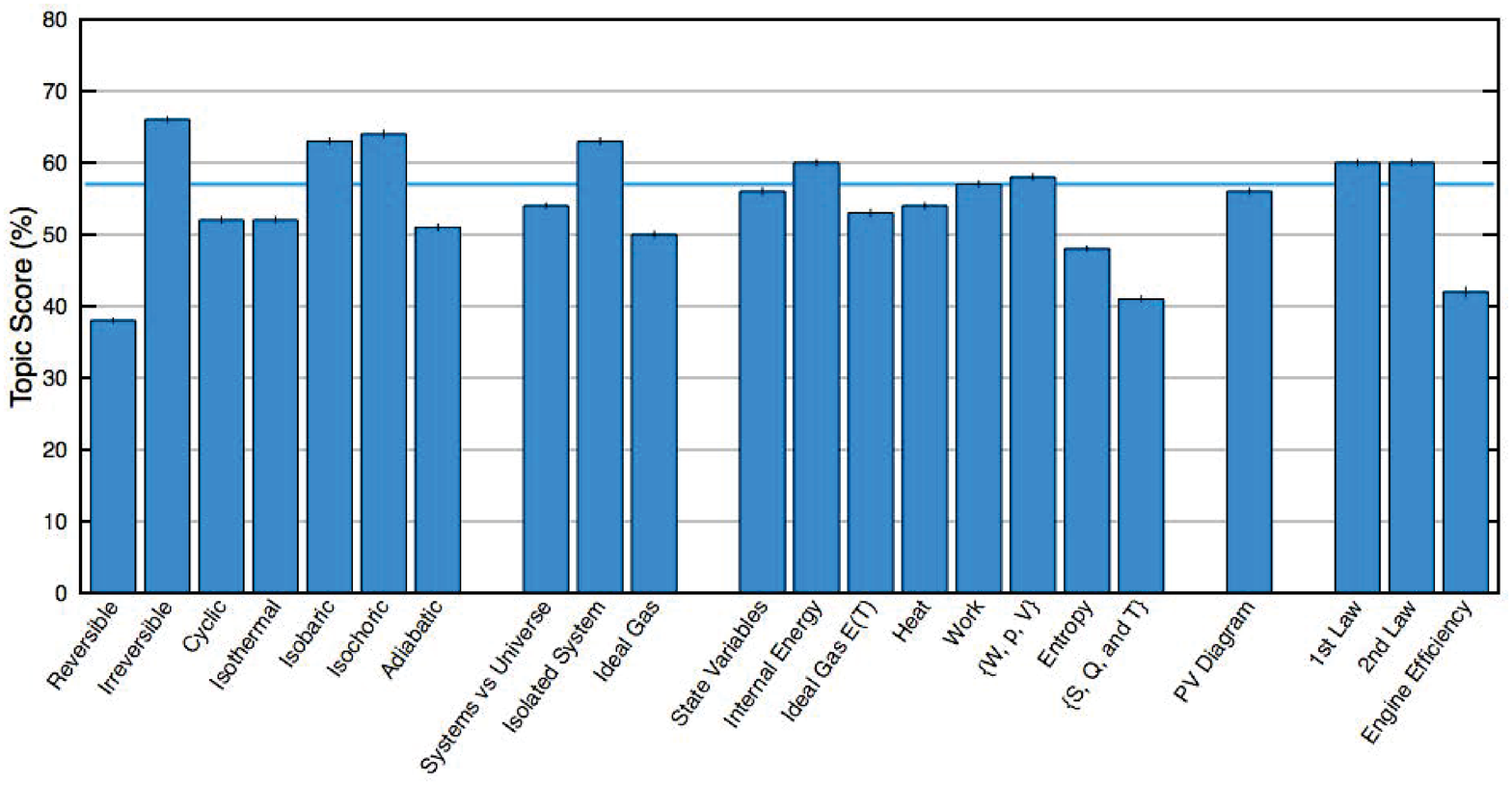}
\caption{Average performance on the STPFaSL instrument by topic for \N=147 upper-level undergraduates and Ph.D. students.  For completeness, error bars on each topic score (very small black vertical lines) indicate the sample error of the mean topic score, \emph{assuming each topic is independent}.  For comparing pairs of topics whose coverage on the STPFaSL instrument is minimally overlapping, e.g., comparison of performance on questions involving Irreversible and {Reversible} processes, an assumption of independence may be appropriate, but otherwise the topics and their errors should not be directly compared within a population.  Another pair of topics for which there is minimal overlap involves the {second law} problems and problems requiring knowledge of the {state variables}. The blue horizontal line shows the average on the entire instrument.}
\label{TopicScoresUpper}

\end{figure}

\begin{figure}[htbp]

\includegraphics[width=\textwidth]{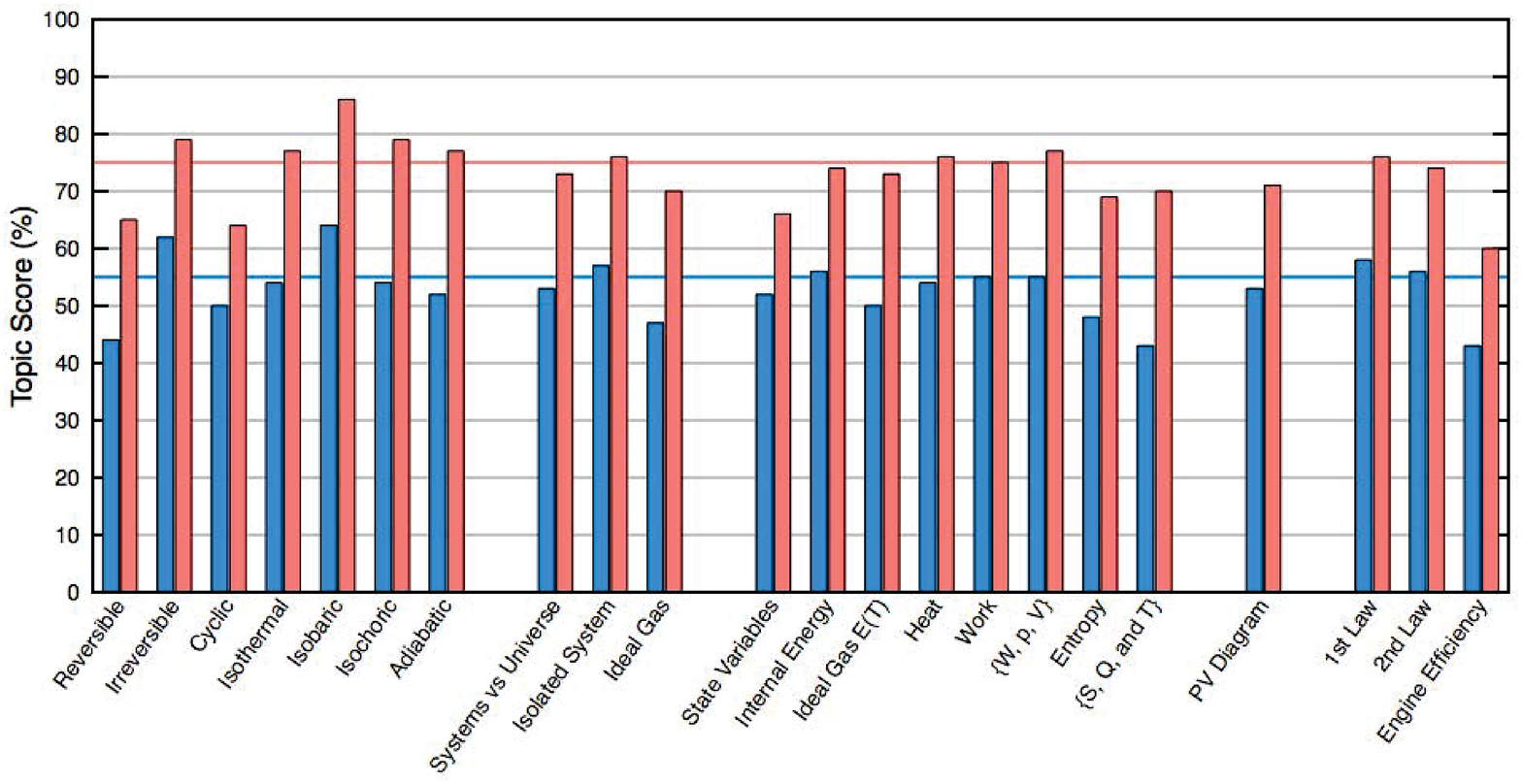}
\caption{Average percentage scores for first-year physics Ph.D. students individually and in pairs on the STPFaSL instrument by topic (for individual Ph.D. students, the number of students \N=45, blue, and for the paired test,  \N=21, red). The blue and red horizontal lines show the averages for those groups on the entire survey instrument.}
\label{TopicScoresGrads}

\end{figure}

Details about student difficulties found using the STPFaSL instrument and comparison with prior studies are beyond the scope of this paper and will be presented elsewhere. However, we note that since the STPFaSL instrument has been administered to a large number of students at six different institutions, quantitative conclusions can be drawn about the prevalence of the many conceptual difficulties students have with these fundamental concepts in thermodynamics (see Table 4 for average performance of each group on each question).   Moreover, Figures \ref{TopicScoresAlgebra} and \ref{TopicScoresCalculus} depict the average percentage scores for students in the algebra-based and calculus-based introductory physics courses, respectively, on the STPFaSL instrument by topic before and after instruction.   
The combined average performance of upper-level undergraduates and physics Ph.D. students in their first semester of the Ph.D. program on various concepts is shown in Figure \ref{TopicScoresUpper} and of Ph.D. students individually vs. those in pairs in Figure \ref{TopicScoresGrads}.  
Some of the conceptual difficulties displayed on the survey instrument include difficulty reasoning with multiple quantities simultaneously, difficulty in systematically applying various constraints (for an isothermal, adiabatic, isochoric, reversible, or irreversible process, isolated system, etc.), difficulty due to oversimplification of the first law and overgeneralization of the second law.  As noted, many of these difficulties were inspired and incorporated in the survey instrument based upon those that have been documented (e.g., see Ref. [38-63]).
Moreover, our findings with this validated survey instrument demonstrate the robustness of the previous findings, e.g., in Ref. [38-63] about student difficulties with these concepts. 
  
 \vspace*{-.2in}
  \section{ Summary}
  \vspace*{-.05in}
We developed and validated and administered a 33-item conceptual multiple-choice survey instrument focusing on thermodynamic processes and the first and second laws at the level covered in introductory physics courses called the Survey of Thermodynamic Processes and First and Second Laws (STPFaSL). The survey instrument uses the common difficulties found in the previous studies and additional data from written responses and interviews as a guide. The concepts related to thermodynamic processes and the first and second laws focusing on topics covered in an introductory physics course turned out to be challenging even for advanced students who were administered the survey instrument for obtaining baseline data and for evaluating content validity. The STPFaSL instrument is designed to measure the effectiveness of traditional and/or research-based approaches for helping introductory students learn these thermodynamics concepts covered in the survey for a group of students. The average individual scores on the survey instrument from traditionally taught classes at various institutions included in this study are low.
We note that the average scores for other conceptual survey instruments for traditionally taught introductory classes are also low, e.g.,  for the BEMA \cite{ding2006evaluating}, the posttest scores for introductory students range from 23\% to 45\%, and for the CSEM \cite{maloney2001surveying}, the scores range from 25\% to 47\%. The low scores even after instruction indicate that the traditional instructional approach using lectures alone is ineffective in helping students learn these concepts. The STPFaSL survey instrument can be used to measure the effectiveness of instruction in these topics using a research-based pedagogy.

\begin{theacknowledgments}

We are extremely indebted to David Meltzer and Mike Loverude for providing very extensive feedback on several versions of the survey instrument. 
We thank all faculty members and students from all six institutions who helped during the development and validation of the survey instrument and/or administered the final version to their classes. We also thank the anonymous reviewers.

\end{theacknowledgments}

\pagebreak

\bibliographystyle{aipproc}   

\pagebreak

\LTcapwidth=\textwidth

\begin{longtable}{r|>{\bfseries}r|rrrr|>{\itshape}l}
\caption{Average percentage scores for each of the five choices for each item on the STPFaSL instrument for each group. Pre or pretest refers to the data before instruction in a particular course in which the survey topics in thermodynamics were covered (as discussed in the text, students in a course may have learned these topics in other courses). Post or posttest refers to data after instruction in relevant concepts in that particular course. Abbreviations for various student groups: Upper (students in junior/senior level thermodynamics and physics Ph.D. students in their first semester of a Ph.D. program who had also only taken the junior/senior level thermodynamics course), calc (students in introductory calculus-based physics courses), Algebra (students in introductory algebra-based physics courses).  The first column shows the percentage of students who answer the item correctly, and the corresponding answer choice.  The four remaining columns list the percentages of incorrect answers (and choices), ranked by frequency. The number of students in each group is the same as in Table 3 except upper-level post and Ph.D. students Ind. are combined into Upper Post.}\label{blocktable}\\

\cline{2-6}
{Problem \#} & Correct (\%) & {$1^{st}$}& {$2^{nd}$} & {$3^{rd}$} & {$4^{th}$} & \normalfont{Level}\\*
\cline{2-6}
\endfirsthead
\cline{2-6}
{Problem \#} & Correct (\%) & {$1^{st}$}& {$2^{nd}$} & {$3^{rd}$} & {$4^{th}$}& \normalfont{Level}\\*
\cline{2-6}
\endhead
\multicolumn{6}{r}{\small\sl continued on next page}\\*
\endfoot
\endlastfoot
1&24 (C)&39 (D)&28 (A)&5 (E)&4 (B)&Upper Post\\*
&28 (C)&32 (A)&29 (D)&8 (B)&3 (E)&Calc Post\\*
&19 (C)&51 (A)&16 (D)&12 (B)&2 (E)&Calc Pre\\*
&29 (C)&36 (D)&29 (A)&4 (B)&2 (E)&Algebra Post\\*
&24 (C)&46 (A)&19 (D)&10 (B)&1 (E)&Algebra Pre\\*
\cline{2-6}
\multicolumn{7}{c}{}\\
\cline{2-6}
2&57 (A)&14 (C)&10 (D)&9 (B)&9 (E)&Upper Post\\*
&35 (A)&24 (C)&16 (D)&13 (E)&11 (B)&Calc Post\\*
&29 (A)&33 (C)&14 (B)&14 (D)&9 (E)&Calc Pre\\*
&39 (A)&20 (C)&15 (D)&14 (E)&13 (B)&Algebra Post\\*
&27 (A)&38 (C)&13 (D)&12 (E)&10 (B)&Algebra Pre\\*
\cline{2-6}
\multicolumn{7}{c}{}\\
\cline{2-6}
3&51 (B)&23 (A)&18 (C)&8 (D)&0 (E)&Upper Post\\*
&28 (B)&38 (C)&22 (A)&11 (D)&1 (E)&Calc Post\\*
&28 (B)&29 (A)&24 (C)&19 (D)&1 (E)&Calc Pre\\*
&29 (B)&33 (C)&19 (D)&18 (A)&1 (E)&Algebra Post\\*
&23 (B)&29 (D)&26 (A)&21 (C)&1 (E)&Algebra Pre\\*
\cline{2-6}
\multicolumn{7}{c}{}\\
\cline{2-6}
4&31 (B)&42 (C)&23 (A)&4 (D)&0 (E)&Upper Post\\*
&28 (B)&36 (C)&26 (A)&9 (D)&1 (E)&Calc Post\\*
&37 (B)&25 (C)&23 (A)&13 (D)&1 (E)&Calc Pre\\*
&32 (B)&43 (C)&12 (A)&11 (D)&1 (E)&Algebra Post\\*
&44 (B)&20 (C)&19 (A)&17 (D)&0 (E)&Algebra Pre\\*
\cline{2-6}
\multicolumn{7}{c}{}\\
\cline{2-6}
5&74 (A)&9 (E)&7 (D)&6 (C)&4 (B)&Upper Post\\*
&68 (A)&18 (E)&6 (C)&4 (D)&4 (B)&Calc Post\\*
&61 (A)&15 (E)&9 (D)&9 (B)&7 (C)&Calc Pre\\*
&50 (A)&27 (E)&9 (D)&8 (C)&5 (B)&Algebra Post\\*
&63 (A)&16 (E)&11 (D)&6 (B)&4 (C)&Algebra Pre\\*
\cline{2-6}
\multicolumn{7}{c}{}\\
\cline{2-6}
6&60 (B)&23 (A)&9 (C)&5 (E)&3 (D)&Upper Post\\*
&27 (B)&37 (C)&21 (A)&9 (D)&5 (E)&Calc Post\\*
&10 (B)&56 (C)&18 (A)&8 (D)&8 (E)&Calc Pre\\*
&9 (B)&47 (C)&28 (A)&10 (D)&6 (E)&Algebra Post\\*
&8 (B)&53 (C)&20 (A)&12 (D)&8 (E)&Algebra Pre\\*
\cline{2-6}
\multicolumn{7}{c}{}\\
\cline{2-6}
7&41 (C)&36 (A)&9 (E)&7 (B)&7 (D)&Upper Post\\*
&53 (C)&22 (A)&13 (E)&6 (B)&6 (D)&Calc Post\\*
&55 (C)&17 (A)&10 (E)&9 (B)&8 (D)&Calc Pre\\*
&43 (C)&22 (E)&21 (A)&7 (B)&7 (D)&Algebra Post\\*
&62 (C)&12 (A)&11 (D)&10 (B)&5 (E)&Algebra Pre\\*
\cline{2-6}
\multicolumn{7}{c}{}\\
\cline{2-6}
8&41 (C)&36 (B)&11 (A)&10 (D)&3 (E)&Upper Post\\*
&21 (C)&33 (A)&27 (B)&12 (D)&7 (E)&Calc Post\\*
&12 (C)&40 (A)&25 (B)&13 (D)&9 (E)&Calc Pre\\*
&10 (C)&37 (A)&24 (B)&19 (D)&10 (E)&Algebra Post\\*
&12 (C)&38 (A)&21 (B)&18 (D)&12 (E)&Algebra Pre\\*
\cline{2-6}
\multicolumn{7}{c}{}\\
\cline{2-6}
9&56 (E)&18 (B)&16 (C)&7 (A)&3 (D)&Upper Post\\*
&40 (E)&23 (C)&17 (B)&11 (D)&9 (A)&Calc Post\\*
&26 (E)&24 (C)&19 (D)&19 (B)&12 (A)&Calc Pre\\*
&38 (E)&21 (C)&17 (B)&15 (D)&10 (A)&Algebra Post\\*
&32 (E)&21 (C)&19 (D)&15 (B)&13 (A)&Algebra Pre\\*
\cline{2-6}
\multicolumn{7}{c}{}\\
\cline{2-6}
10&53 (E)&19 (D)&11 (B)&11 (A)&6 (C)&Upper Post\\*
&23 (E)&30 (B)&23 (A)&14 (D)&10 (C)&Calc Post\\*
&15 (E)&40 (B)&27 (A)&9 (C)&9 (D)&Calc Pre\\*
&23 (E)&36 (B)&27 (A)&8 (D)&6 (C)&Algebra Post\\*
&17 (E)&41 (B)&28 (A)&11 (C)&3 (D)&Algebra Pre\\*
\cline{2-6}
\multicolumn{7}{c}{}\\
\cline{2-6}
11&80 (C)&9 (A)&4 (E)&4 (D)&3 (B)&Upper Post\\*
&69 (C)&14 (A)&8 (B)&6 (D)&2 (E)&Calc Post\\*
&57 (C)&17 (A)&11 (B)&11 (D)&4 (E)&Calc Pre\\*
&65 (C)&19 (A)&10 (B)&3 (D)&2 (E)&Algebra Post\\*
&50 (C)&26 (A)&11 (D)&9 (B)&4 (E)&Algebra Pre\\*
\cline{2-6}
\multicolumn{7}{c}{}\\
\cline{2-6}
12&62 (D)&17 (A)&13 (E)&6 (C)&2 (B)&Upper Post\\*
&31 (D)&40 (A)&15 (E)&10 (C)&5 (B)&Calc Post\\*
&24 (D)&44 (A)&13 (E)&13 (C)&6 (B)&Calc Pre\\*
&32 (D)&41 (A)&12 (C)&12 (E)&3 (B)&Algebra Post\\*
&29 (D)&39 (A)&17 (E)&11 (C)&4 (B)&Algebra Pre\\*
\cline{2-6}
\multicolumn{7}{c}{}\\
\cline{2-6}
13&74 (C)&14 (E)&9 (B)&3 (A)&0 (D)&Upper Post\\*
&43 (C)&24 (E)&24 (B)&5 (D)&4 (A)&Calc Post\\*
&36 (C)&30 (E)&23 (B)&7 (D)&5 (A)&Calc Pre\\*
&61 (C)&18 (B)&14 (E)&4 (A)&3 (D)&Algebra Post\\*
&43 (C)&24 (B)&19 (E)&10 (D)&4 (A)&Algebra Pre\\*
\cline{2-6}
\multicolumn{7}{c}{}\\
\cline{2-6}
14&55 (E)&18 (C)&16 (D)&8 (A)&3 (B)&Upper Post\\*
&30 (E)&32 (C)&17 (A)&15 (D)&6 (B)&Calc Post\\*
&21 (E)&28 (C)&20 (D)&17 (A)&14 (B)&Calc Pre\\*
&28 (E)&31 (C)&19 (D)&19 (A)&3 (B)&Algebra Post\\*
&20 (E)&28 (D)&24 (C)&16 (A)&12 (B)&Algebra Pre\\*
\cline{2-6}
\multicolumn{7}{c}{}\\
\cline{2-6}
15&40 (E)&29 (D)&22 (A)&6 (C)&4 (B)&Upper Post\\*
&22 (E)&43 (A)&17 (D)&11 (C)&8 (B)&Calc Post\\*
&11 (E)&43 (A)&18 (C)&14 (D)&13 (B)&Calc Pre\\*
&11 (E)&50 (A)&14 (D)&13 (B)&12 (C)&Algebra Post\\*
&9 (E)&53 (A)&15 (D)&12 (C)&11 (B)&Algebra Pre\\*
\cline{2-6}
\multicolumn{7}{c}{}\\
\cline{2-6}
16&58 (D)&22 (E)&15 (C)&4 (B)&1 (A)&Upper Post\\*
&27 (D)&27 (C)&27 (E)&14 (B)&5 (A)&Calc Post\\*
&19 (D)&33 (C)&21 (E)&17 (B)&10 (A)&Calc Pre\\*
&27 (D)&40 (E)&19 (C)&9 (B)&5 (A)&Algebra Post\\*
&16 (D)&36 (C)&24 (E)&12 (B)&11 (A)&Algebra Pre\\*
\cline{2-6}
\multicolumn{7}{c}{}\\
\cline{2-6}
17&72 (E)&14 (C)&7 (B)&4 (A)&3 (D)&Upper Post\\*
&58 (E)&19 (C)&10 (A)&8 (B)&5 (D)&Calc Post\\*
&50 (E)&19 (C)&12 (B)&11 (A)&9 (D)&Calc Pre\\*
&48 (E)&21 (C)&14 (A)&10 (B)&7 (D)&Algebra Post\\*
&55 (E)&20 (C)&9 (A)&9 (B)&7 (D)&Algebra Pre\\*
\cline{2-6}
\multicolumn{7}{c}{}\\
\cline{2-6}
18&75 (B)&9 (C)&6 (D)&6 (E)&3 (A)&Upper Post\\*
&37 (B)&28 (C)&16 (E)&11 (D)&8 (A)&Calc Post\\*
&20 (B)&34 (C)&16 (E)&15 (A)&15 (D)&Calc Pre\\*
&46 (B)&21 (E)&18 (C)&9 (D)&6 (A)&Algebra Post\\*
&37 (B)&23 (C)&15 (D)&14 (E)&12 (A)&Algebra Pre\\*
\cline{2-6}
\multicolumn{7}{c}{}\\
\cline{2-6}
19&28 (E)&34 (B)&19 (C)&16 (D)&4 (A)&Upper Post\\*
&25 (E)&24 (B)&19 (C)&17 (D)&15 (A)&Calc Post\\*
&18 (E)&26 (B)&20 (D)&19 (C)&17 (A)&Calc Pre\\*
&37 (E)&25 (D)&18 (B)&10 (A)&10 (C)&Algebra Post\\*
&21 (E)&27 (D)&22 (B)&18 (C)&11 (A)&Algebra Pre\\*
\cline{2-6}
\multicolumn{7}{c}{}\\
\cline{2-6}
20&51 (A)&20 (C)&16 (D)&10 (B)&3 (E)&Upper Post\\*
&31 (A)&30 (C)&16 (D)&14 (B)&9 (E)&Calc Post\\*
&23 (A)&26 (C)&22 (D)&17 (B)&12 (E)&Calc Pre\\*
&44 (A)&22 (C)&14 (B)&11 (D)&9 (E)&Algebra Post\\*
&23 (A)&24 (C)&20 (B)&18 (D)&16 (E)&Algebra Pre\\*
\cline{2-6}
\multicolumn{7}{c}{}\\
\cline{2-6}
21&70 (B)&12 (E)&10 (D)&5 (C)&2 (A)&Upper Post\\*
&49 (B)&16 (D)&16 (E)&9 (A)&9 (C)&Calc Post\\*
&35 (B)&20 (C)&19 (D)&13 (A)&13 (E)&Calc Pre\\*
&44 (B)&21 (E)&15 (D)&12 (C)&9 (A)&Algebra Post\\*
&37 (B)&19 (D)&18 (C)&16 (E)&10 (A)&Algebra Pre\\*
\cline{2-6}
\multicolumn{7}{c}{}\\
\cline{2-6}
22&78 (D)&16 (B)&4 (C)&3 (A)&0 (E)&Upper Post\\*
&52 (D)&16 (B)&12 (E)&12 (C)&7 (A)&Calc Post\\*
&44 (D)&19 (E)&17 (B)&12 (C)&8 (A)&Calc Pre\\*
&49 (D)&16 (B)&13 (C)&11 (A)&11 (E)&Algebra Post\\*
&51 (D)&18 (E)&16 (B)&10 (C)&6 (A)&Algebra Pre\\*
\cline{2-6}
\multicolumn{7}{c}{}\\
\cline{2-6}
23&62 (A)&22 (C)&9 (D)&5 (B)&2 (E)&Upper Post\\*
&52 (A)&20 (C)&14 (D)&11 (B)&2 (E)&Calc Post\\*
&43 (A)&21 (C)&18 (B)&12 (D)&5 (E)&Calc Pre\\*
&46 (A)&20 (C)&18 (D)&14 (B)&1 (E)&Algebra Post\\*
&40 (A)&25 (B)&20 (C)&11 (D)&3 (E)&Algebra Pre\\*
\cline{2-6}
\multicolumn{7}{c}{}\\
\cline{2-6}
24&64 (D)&15 (A)&14 (E)&5 (C)&2 (B)&Upper Post\\*
&32 (D)&42 (A)&12 (E)&9 (C)&6 (B)&Calc Post\\*
&24 (D)&39 (A)&13 (C)&12 (E)&11 (B)&Calc Pre\\*
&31 (D)&41 (A)&11 (E)&9 (C)&8 (B)&Algebra Post\\*
&30 (D)&43 (A)&11 (C)&9 (B)&7 (E)&Algebra Pre\\*
\cline{2-6}
\multicolumn{7}{c}{}\\
\cline{2-6}
25&69 (C)&20 (E)&9 (B)&2 (D)&0 (A)&Upper Post\\*
&38 (C)&26 (B)&23 (E)&7 (D)&6 (A)&Calc Post\\*
&29 (C)&28 (B)&25 (E)&12 (D)&5 (A)&Calc Pre\\*
&57 (C)&18 (E)&17 (B)&5 (D)&3 (A)&Algebra Post\\*
&35 (C)&21 (B)&21 (E)&15 (D)&8 (A)&Algebra Pre\\*
\cline{2-6}
\multicolumn{7}{c}{}\\
\cline{2-6}
26&65 (E)&12 (D)&11 (B)&10 (C)&2 (A)&Upper Post\\*
&41 (E)&19 (B)&18 (C)&15 (D)&7 (A)&Calc Post\\*
&17 (E)&25 (B)&23 (D)&22 (C)&13 (A)&Calc Pre\\*
&31 (E)&26 (B)&20 (D)&14 (C)&9 (A)&Algebra Post\\*
&17 (E)&27 (D)&27 (B)&18 (C)&11 (A)&Algebra Pre\\*
\cline{2-6}
\multicolumn{7}{c}{}\\
\cline{2-6}
27&74 (D)&10 (C)&6 (E)&6 (B)&4 (A)&Upper Post\\*
&34 (D)&31 (C)&13 (E)&11 (B)&10 (A)&Calc Post\\*
&25 (D)&27 (C)&21 (B)&15 (E)&11 (A)&Calc Pre\\*
&35 (D)&29 (C)&16 (B)&11 (E)&9 (A)&Algebra Post\\*
&28 (D)&25 (C)&25 (B)&14 (A)&8 (E)&Algebra Pre\\*
\cline{2-6}
\multicolumn{7}{c}{}\\
\cline{2-6}
28&14 (E)&26 (C)&26 (D)&20 (B)&14 (A)&Upper Post\\*
&14 (E)&27 (D)&27 (B)&17 (C)&16 (A)&Calc Post\\*
&11 (E)&25 (B)&23 (D)&22 (A)&18 (C)&Calc Pre\\*
&14 (E)&30 (D)&28 (B)&14 (A)&14 (C)&Algebra Post\\*
&9 (E)&33 (D)&23 (A)&18 (B)&18 (C)&Algebra Pre\\*
\cline{2-6}
\multicolumn{7}{c}{}\\
\cline{2-6}
29&68 (A)&15 (B)&9 (D)&7 (C)&1 (E)&Upper Post\\*
&43 (A)&24 (B)&17 (C)&8 (D)&7 (E)&Calc Post\\*
&35 (A)&23 (B)&20 (C)&15 (D)&7 (E)&Calc Pre\\*
&57 (A)&16 (B)&13 (C)&10 (D)&4 (E)&Algebra Post\\*
&40 (A)&21 (B)&17 (D)&15 (C)&7 (E)&Algebra Pre\\*
\cline{2-6}
\multicolumn{7}{c}{}\\
\cline{2-6}
30&37 (D)&24 (A)&22 (C)&11 (B)&5 (E)&Upper Post\\*
&19 (D)&36 (C)&23 (B)&19 (A)&4 (E)&Calc Post\\*
&17 (D)&33 (C)&25 (B)&18 (A)&8 (E)&Calc Pre\\*
&14 (D)&47 (C)&22 (B)&9 (A)&8 (E)&Algebra Post\\*
&11 (D)&42 (C)&21 (A)&18 (B)&8 (E)&Algebra Pre\\*
\cline{2-6}
\multicolumn{7}{c}{}\\
\cline{2-6}
31&57 (A)&24 (C)&11 (B)&6 (D)&2 (E)&Upper Post\\*
&42 (A)&20 (B)&18 (C)&16 (D)&4 (E)&Calc Post\\*
&36 (A)&21 (C)&20 (B)&17 (D)&5 (E)&Calc Pre\\*
&46 (A)&21 (D)&16 (C)&14 (B)&2 (E)&Algebra Post\\*
&35 (A)&25 (B)&21 (D)&16 (C)&3 (E)&Algebra Pre\\*
\cline{2-6}
\multicolumn{7}{c}{}\\
\cline{2-6}
32&62 (C)&21 (A)&8 (D)&6 (B)&3 (E)&Upper Post\\*
&37 (C)&23 (A)&21 (D)&13 (B)&7 (E)&Calc Post\\*
&33 (C)&23 (D)&21 (A)&15 (B)&8 (E)&Calc Pre\\*
&40 (C)&22 (A)&18 (D)&15 (B)&5 (E)&Algebra Post\\*
&27 (C)&31 (A)&16 (D)&13 (B)&12 (E)&Algebra Pre\\*
\cline{2-6}
\multicolumn{7}{c}{}\\
\cline{2-6}
33&75 (C)&12 (E)&6 (B)&4 (D)&2 (A)&Upper Post\\*
&43 (C)&22 (B)&19 (E)&9 (D)&7 (A)&Calc Post\\*
&32 (C)&24 (B)&20 (E)&13 (A)&10 (D)&Calc Pre\\*
&49 (C)&18 (E)&15 (B)&12 (D)&6 (A)&Algebra Post\\*
&25 (C)&23 (A)&19 (B)&18 (E)&16 (D)&Algebra Pre\\*
\cline{2-6}
\end{longtable}

\clearpage

\renewcommand{\arraystretch}{1.05}
\setlength{\bigstrutjot}{6pt}

\pagebreak
\clearpage


\end{document}